\documentclass[11pt]{article}
\usepackage{amsmath}
\usepackage{graphicx}
\usepackage{centernot}
\usepackage{natbib}
\usepackage{url} 
\usepackage[shortlabels]{enumitem}
\usepackage{latexsym, epsfig, amssymb, amsthm, mathrsfs, bbm, enumitem}
\usepackage[OT1]{fontenc}
\usepackage{xcolor}
\usepackage[colorlinks,citecolor=blue,linkcolor=blue,urlcolor=blue]{hyperref}
\usepackage{multirow,multicol,booktabs}
\usepackage{anysize}

\marginsize{1.1in}{1.1in}{0.55in}{0.8in} %

\makeatletter

\usepackage[ruled]{algorithm2e}
\usepackage{tikz}
\usepackage{float}
\usepackage{booktabs}
\usepackage{adjustbox}

\newcommand{\bdelta}{\mbox{\boldmath $\delta$}}
\newcommand{\bSigma}{\mbox{\boldmath $\Sigma$}}

\newtheorem{definition}{Definition}
\newtheorem{theorem}{Theorem}
\newtheorem{lemma}{Lemma}
\newtheorem{condition}{Condition}

\def\X{{\bf X}}

\def\bx{{\bf x}}
\def\y{{\bf y}}

\def\u{{\bf u}}
\def\v{{\bf v}}
\def\bz{{\bf z}}

\def\bb{{\bf b}}
\def\bs{{\bf s}}

\def\0{{\bf 0}}
\newcommand{\bSig}{\mbox{\boldmath $\Sigma$}}

\newcommand{\bbeta}{\mbox{\boldmath $\beta$}}

\newcommand{\balpha}{\mbox{\boldmath $\alpha$}}
\newcommand{\bgamma}{\mbox{\boldmath $\gamma$}}

\DeclareMathOperator*{\argmin}{arg\,min}

\begin{document}

\title{CoxKnockoff: Controlled Feature Selection for the Cox Model Using Knockoffs%
\thanks{ 
Daoji Li is Assistant Professor, Department of Information Systems and Decision Sciences, California State
University, Fullerton, CA, 92831, United States (E-mail: dali@fullerton.edu). %
Jinzhao Yu is Ph.D. candidate, School of Statistics and Mathematics, Zhongnan University of Economics and Law, Wuhan, 430073, China (E-mail: yjz@stu.zuel.edu.cn). %
Hui Zhao is Professor, School of Statistics and Mathematics, Zhongnan University of Economics and Law, Wuhan, 430073, China (E-mail: hzhao@zuel.edu.cn). %
Li was supported by the 2022-2023 RSCA Award Program at California State University, Fullerton, United States. Zhao was partially supported by National Natural Science Foundation of China (Grant 12171483). 
}
\date{
\medskip
\medskip
This version has been accepted for publication in \href{https://onlinelibrary.wiley.com/journal/20491573}{\textit{Stat}} on July 11, 2023;\\
see \url{https://doi.org/10.1002/sta4.607} for the final published version
}
\author{Daoji Li$^1$, Jinzhao Yu$^2$ and Hui Zhao$^2$
\medskip\\
$^1$California State University, Fullerton
\\ $^2$Zhongnan University of Economics and Law
\\
} %
}

\maketitle

\begin{abstract}	
Although there is a huge literature on feature selection for the Cox model, none of the existing approaches can control the false discovery rate (FDR) unless the sample size tends to infinity.
In addition, there is no formal power analysis of the knockoffs framework for survival data in the literature. To address those issues, in this paper, we propose a novel controlled feature selection approach using knockoffs for the Cox model. 
We establish that the proposed method enjoys the FDR control in finite samples regardless of the number of covariates.
Moreover, under mild regularity conditions, we also show that the power of our method is asymptotically one as sample size tends to infinity. To the best of our knowledge, this is the first formal theoretical result on the power for the knockoffs procedure  
in the survival setting. Simulation studies confirm that our method has appealing finite-sample performance with desired FDR control and high power. 
We further demonstrate the performance of our method through a real data example.
\end{abstract}
	
\textit{Running title}: CoxKnockoff


\newpage
\section{Introduction}\label{sec:Intro}

The Cox model~\citep{cox1972regression}, also known as the proportional hazards model, is one of the most popular regression models for analysis of survival data. With recent technological advances, survival data with a large number of explanatory features are ubiquitous in biomedical, social, behavioral, and epidemiological studies. When a large number of explanatory features are available, feature selection is fundamental in analysis of survival data because it can prevent overfitting, enhance model interpretation, and improve statistical accuracy. The objective of feature selection is to identify the truly relevant variables that contribute to the effect of the response variable from a large number of features.

Many regularization methods have been proposed for feature selection in the Cox model; see, for example, \cite{tibshirani1997lasso}, \cite{fan2002variable}, 
\cite{zhang2007adaptive}, \cite{zou2008note}, \cite{antoniadis2010dantzig},
\cite{bradic2011regularization},
\cite{huang2013oracle},
and references therein. However, none of the existing feature selection approaches for the Cox model can control the false discovery rate (FDR) unless the sample size tends to infinity. FDR was formally introduced by~\cite{benjamini1995controlling} and is defined as the expectation of the false discovery proportion.
To achieve FDR control, \cite{benjamini1995controlling} proposed the so-called B-H procedure by resorting to the $p$-values for large-scale multiple testing, which  
can guarantee exact FDR control when all the $p$-values are independent. \cite{benjamini2001control} extended the B-H procedure to deal with dependent $p$-values. See also \cite{storey2002direct}, \cite{abramovich2006adapting},  \cite{fan2012estimating}, and \cite{sarkar2022adjusting} for other FDR control procedures along this line.  

Another popular class of FDR control methods is based on the knockoffs framework, which does not use $p$-values and can achieve finite-sample FDR control by constructing a set of knockoff features that mimic the dependence structure of the original ones. The first knockoffs procedure, named fixed-X knockoffs in the literature, was proposed by \cite{barber2015controlling} for 
low-dimensional Gaussian linear models with fixed design and extended to high-dimensional linear models by \cite{barber2019knockoff}.
Recently, \cite{candes2018panning} introduced the model-X knockoffs framework, which removes the linearity assumption in fixed-X knockoffs, works for any sample size and any number of covariates,
and guarantees exact FDR control in finite samples for arbitrary dimensions without knowing the underlying true relationship between the response and features. The robustness of 
model-X knockoffs  
was further studied in \cite{barber2020robust} and \cite{fan2020rank}. 
The knockoffs framework has been used to control FDR under 
different  
models and problems, such as latent factor models~\citep{fan2020ipad}, Gaussian graphical models~\citep{zheng2018recovering, li2021ggm, zhou2022reproducible}, and feature screening~\citep{liu2022model}.

However, there is no existing result on finite-sample FDR control for the Cox model in the literature. 
Moreover, 
formal power analysis
of those existing knockoffs-based feature selection approaches is generally lacking, 
except for~\cite{fan2020rank} and~\cite{weinstein2020power} (both for linear models) and~\cite{fan2020ipad} (for latent factor models). There is no formal power analysis
of the knockoffs framework for censored data.
As pointed out by \cite{fan2020rank}, the power analysis for the knockoffs framework is necessary and nontrivial. Similar to Type I and Type II errors in hypothesis testing, the FDR and power are two sides of the same coin.

In this paper, we propose a novel feature selection method with guaranteed FDR control and high power for the Cox model.
Our proposed method, named CoxKnockoff, combines the strengths of model-X knockoffs and penalized partial log-likelihood estimation. We have shown that CoxKnockoff can control the FDR at a target level in finite samples regardless of the number of covariates.
Moreover, under mild regularity conditions, we establish that the power of CoxKnockoff is asymptotically one as the sample size tends to infinity. 
To the best of our knowledge, this is the first formal theoretical result on the power for the knockoffs procedure in the survival setting.

The rest of the article is organized as follows. Section~\ref{sec: method} introduces the model setting, reviews model-X knockoffs, and presents our procedure CoxKnockoff.  
Section~\ref{sec: theory} establishes the theoretical guarantees of CoxKnockoff on
the FDR and power.
Sections~\ref{sec: simulations} and~\ref{sec: RealDataAnalysis} 
provide several simulation and real
data examples to demonstrate the finite-sample performance and the advantages of 
CoxKnockoff, respectively. 
All the proofs and technical details are given in the Supporting Information.

\section{Controlled feature selection in the Cox model}\label{sec: method}

\subsection{Model Setting}\label{sec: model-setting}
Let $T$ and $\bx = (X_{1}, \ldots, X_{p})^{\top}$ be the survival time and the $p$-dimensional covariate vector, respectively. Since the survival time $T$ is generally subject to right censoring, one can only observe $(Y, \bx^{\top}, \delta)$, where $Y=\min\{T,\, U\}$ is the  observed time, $U$ is the censoring time, and $\delta=I(T\leq U)$ is the censoring indicator. 
In this paper, we assume that $T$ and
$U$ are conditionally independent given $\bx$ and consider the following Cox model
\begin{align*}
h(t|\bx) = h_0(t)\exp(\bbeta^{\top}\bx),
\end{align*}
where $h(t|\bx)$ is the hazard at time $t$ given covariate vector $\bx$, 
$h_0(t)$ is an unspecified baseline hazard function, and $\bbeta=(\beta_{1}, \ldots, \beta_{p})^{\top}$ is a $p$-vector of unknown regression coefficients. 
Suppose that the observed data consist of $(Y_i, \bx_i^{\top}, \delta_i), i=1, \ldots, n$, which are independent copies of $(Y, \bx^{\top}, \delta)$.
To simplify our presentation, assume that there are no ties in the observed times $Y_i$'s, otherwise we can use  the technique in \cite{breslow1974covariance}. Then the partial log-likelihood
of the observed data is 
\begin{equation*}
\ell(\bbeta)
=n^{-1}\sum_{i=1}^{n}\delta_i
\Big\{\bbeta^{\top}\bx_i-\log\Big[\sum\limits_{j=1}^{n}I(Y_j\geq Y_i) \exp(\bbeta^{\top}\bx_j)\Big] \Big\}.
\end{equation*}

Denote by $\bbeta_0=(\beta_{01}, \ldots, \beta_{0p})^{\top}\in\mathbb{R}^p$ the true coefficient vector.
Let $\mathcal{S}_0=\{1\leq j\leq p:\beta_{0j}\neq 0\}$ denote the index set of covariates that are truly relevant to the hazard function and $\mathcal{S}_0^c=\{1, \ldots, p\}\setminus\mathcal{S}_0$ be the index set of the noise covariates.  
The FDR of a feature selection procedure with an identified set $\widehat{\mathcal{S}}_0$ of relevant covariates is 
\begin{equation*}\label{FDRdf}
\mathrm{FDR} = \mathbb{E}[\mbox{FDP}]\,\,\mbox{with}\,\,
\mbox{FDP} = \frac{|\widehat{\mathcal{S}}_0 \cap \mathcal{S}_0^c|}{|\widehat{\mathcal{S}}_0|\vee 1},
\end{equation*}
where $|\cdot|$ denotes the cardinality of a given set, the expectation is taken with respect to the randomness in $\widehat{\mathcal{S}}_0$, and we use the notation $a \vee b=\max\{a, b\}$. A modified version of FDR (mFDR) is defined as
\begin{equation*}\label{mFDRdf}
\mathrm{mFDR} = \mathbb{E} \left[\frac{|\widehat{\mathcal{S}}_0 \cap \mathcal{S}_0^c|}{|\widehat{\mathcal{S}}_0|+q^{-1}}\right],
\end{equation*}
where $q\in(0, 1)$ is the target FDR level. 
Obviously, FDR is more conservative than mFDR since controlling the FDR naturally results in the control of mFDR. It is also warranted to consider another important performance measure, power, to study the capability of a feature selection procedure in discovering the truly relevant covariates. More precisely, power is defined as the expectation of the true discovery proportion, that is,
\begin{equation*}\label{power}
\mathrm{power}(\widehat{\mathcal{S}}_0 ) = \mathbb{E} \left[\frac{|\widehat{\mathcal{S}}_0 \cap \mathcal{S}_0|}{|\mathcal{S}_0|}\right].
\end{equation*}
A desirable reproducible feature selection approach should be able to control the FDR at a pre-determined target level and achieve high power.

Our goal is to develop a feature selection method that can not only identify the truly relevant features, but also control the FDR within a target level.  
To achieve this goal, we will use the model-X knockoffs framework introduced in \cite{candes2018panning}. The main idea of our method is to 
add knockoff variables, which are constructed from the original covariates without using any information of the observed times, in penalized partial log-likelihood estimation.
Knockoff variables will play a
key role in controlled feature selection by serving as negative controls that allow us to estimate and limit the number of false positives.  For completeness, we briefly review the model-X knockoffs below.

\subsection{Model-X Knockoffs}

\begin{definition}\label{def1}
	\emph{(Definition 2 in \cite{candes2018panning})}. The model-X knockoffs for a set of random variables $\bx =(X_1,\cdots,X_p)^{\top}$ are a new set of random variables $\widetilde{\bx}=(\widetilde{X}_1, \cdots, \widetilde{X}_p)^{\top}$ satisfying
	two properties: 
	(a) $(\bx^{\top}, \widetilde{\bx}^{\top})_{\mathrm{swap}(\mathcal{S})}\overset{d}{=}(\bx^{\top}, \widetilde{\bx}^{\top})$ for any subset $\mathcal{S}\subset \{1,\cdots,p\}$,
	where $\mathrm{swap}(\mathcal{S})$ means
	swapping components $X_j$ and $\widetilde{X}_j$ for each $j \in \mathcal{S}$ and $\overset{d}{=}$ denotes 
	being identical in distribution; 
	(b) conditional on $\bx$, the response is independent of the knockoffs vector $\widetilde{\bx}$.
\end{definition}

Property (a) requires
the original and knockoff variables to be pairwise exchangeable while property (b) 
is satisfied when $\widetilde{\bx}$ is constructed without using any information of the response variable. The construction of knockoff variables is cheap because it does not require collecting any
new data. When the distribution of $\bx$ is known, one can use 
the SCIP algorithm proposed in \cite{candes2018panning} to construct the exact  knockoff variables $\widetilde{\bx}$.
When the distribution of $\bx$ is unknown,  
\cite{candes2018panning} suggested  
to construct second-order knockoff variables $\widetilde{\bx}$ such that $(\bx,\widetilde{\bx})$ is pairwise exchangeable with respect to the first two moments. Equivalently, $\widetilde{\bx}$ is the second-order knockoff copy of $\bx$ if 
$\mathbb{E}(\bx)=\mathbb{E}(\widetilde{\bx})$ and 
\begin{equation}\label{eq: cov}
\mathrm{cov}(\bx,\widetilde{\bx})=
\begin{pmatrix}
&\bSig  &\bSig - {\rm{diag}}\{\bs\} \\
&\bSig - {\rm{diag}}\{\bs\} &\bSig
\end{pmatrix},
\end{equation}
where $\bSig\in \mathbb{R}^{p\times p}$ is the covariance matrix of $\bx$
and
$\bs=(s_1, \cdots, s_p)^T\in \mathbb{R}^{p} $ is a vector such that the covariance matrix $\mathrm{cov}(\bx,\widetilde{\bx})$ is positive semi-definite. 
Assume that each  covariate $X_j$ has been rescaled to have mean zero and variance one.
Then the vector $\bs$
can be chosen using two methods introduced in \cite{barber2015controlling} and \cite{candes2018panning}. The first one is the equicorrelated construction, which takes
\begin{align*}
s_j = 2\lambda_{\min}(\bSig)\wedge 1\,\,\,\mbox{for all}\,\,\,j=1, \ldots, p,
\end{align*}
where we use the notation $a \wedge b=\min\{a, b\}$. The second one is the semidefinite program construction, which solves the convex problem
\begin{align*}
\mbox{minimize}\sum_{j=1}^p|1-s_j|\,\,\,\mbox{subject to}
\,\,\,s_j\geq 0\,\,\, \mbox{and}\,\,\, 2\bSig\succeq \mbox{diag}\{\bs\},
\end{align*}
where $A\succeq B$ means that $A-B$ is positive-semidefinite.
See \cite{candes2018panning} for full details on 
the construction of model-X knockoff variables.

\subsection{Our Methodology: CoxKnockoff}

Now we are ready to introduce our method, CoxKnockoff, for controlled feature selection in the Cox model using knockoffs. It consists of three steps, which are described below.

{\bf Step 1. Construct knockoff variables}.  The first step of our approach is to construct knockoffs variables using original covariates only. Given the original covariates $\bx_i, i=1, \ldots, n$, here we employ the aforementioned procedures in \cite{candes2018panning} to generate their 
knockoff copies 
$\widetilde{\bx}_i$.

{\bf Step 2. Calculate statistics}.
After obtaining knockoff copies $\widetilde{\bx}_1, \ldots, \widetilde{\bx}_n$, we can construct the
knockoff statistics to find the relevant variables. To do so, 
we consider the following penalized estimator from the partial log-likelihood with the original covariates and knockoff variables
\begin{align}\label{eq: hat-b}
\widehat{\bb}(\lambda)
=\argmin_{\bb\in\mathbb{R}^{2p}}\left\{-\widetilde{\ell}(\bb)+\lambda\|\bb\|_1\right\},
\end{align}
where  $\widetilde{\ell}(\bb)=n^{-1}\sum\limits_{i=1}^{n}\delta_i
\Big\{\bb^{\top}\bz_i-\log\Big[\sum\limits_{j=1}^{n}I(Y_j\geq Y_i) \exp(\bb^{\top}\bz_j)\Big] \Big\}$ with $\bz_i=(\bx_i^{\top}, \widetilde{\bx}_i^{\top})^{\top}\in\mathbb{R}^{2p}$ for each $i=1. \ldots, n$, $\lambda>0$ is the regularization parameter, and $\|\cdot\|_1$ is the $L_1$ norm of a vector.
Observe that the first $p$ components of $\widehat{\bb}(\lambda)$ are the coefficients of the original
variables and the last $p$ components are for the knockoff variables. Therefore, we use the following Lasso coefficient difference (LCD) 
\begin{align}\label{eq: Statistic-Wj}
W_j = |\widehat{b}_j(\lambda)| - |\widehat{b}_{j+p}(\lambda)|
\end{align}
as our knockoff statistic, where $\widehat{b}_j(\lambda)$ and $\widehat{b}_{j+p}(\lambda)$ are the $j$th and $(j+p)$th components of $\widehat{\bb}(\lambda)$, respectively.
It follows from property (b) in Definition \ref{def1} that all knockoff variables are just noise features. Therefore,  
intuitively, if the original covariate $X_j$ is truly relevant, $W_j$ is likely to have a large positive value.  On the other hand, the magnitude of $W_j$ is expected to be small if $X_j$ is not relevant. In other words, knockoff statistics $W_j$'s can be used to measure the importance of original covariates.

{\bf Step 3. 
	Estimate the set of relevant covariates $\mathcal{S}_0$}. The final step of our procedure is to estimate the set of relevant covariates $\mathcal{S}_0$ based on the knockoff statistics $W_j$, $j=1, \ldots, p$. 
To this end, we can select those covariates whose $W_j$ are greater than or equal to some positive threshold.  In order to achieve FDR control, 
we follow the ideas
in \cite{barber2015controlling} and \cite{candes2018panning} to obtain the  threshold. More precisely, for a target FDR level $0<q<1$ specified by the user, the set of relevant covariates $\mathcal{S}_0$ is estimated as 
\begin{align}\label{eq: S0_hat}
\widehat{\mathcal{S}}_0 = \{1\leq j\leq p: W_j\geq t\}\,\,\,\mbox{with}\,\,t=\tau \,\,\,\mbox{or}\,\,\, \tau_{+}, 
\end{align}
where $\tau$ is the knockoff threshold 
\begin{align}\label{eq: tau}
\tau = \min\left\{t\in \mathcal{W}: \frac{|\{j: W_j \leq -t\}|}{|\{j: W_j \geq t\}|\vee 1}\leq q\right\}
\end{align}
and $\tau_{+}$ is the knockoff+ threshold 
\begin{align}\label{eq: tau+}
\tau_+ = \min\left\{t\in \mathcal{W}: \frac{|\{j: W_j \leq -t\}|+1}{|\{j: W_j \geq t\}|\vee 1}\leq q\right\}.
\end{align}
Here $\mathcal{W}=\{|W_j|: 1\leq j\leq p\}\setminus\{0\}$ and $\widehat{\mathcal{S}}_0$ is defined as an empty set if the threshold $\tau=+\infty$ or $\tau_{+}=+\infty$. As shown later in our Theorem~\ref{Theorem1}, the knockoff threshold can control the mFDR while the knockoff+ threshold controls the exact FDR 
in finite samples regardless of the number of covariates $p$.

\section{Theoretical Properties}\label{sec: theory}

In this section, we investigate the theoretical properties of CoxKnockoff.  
We first show that the FDR control is guaranteed for
CoxKnockoff at any target level 
for any sample size $n$ and any number of covariates $p$. 
Then we 
establish that, under mild regularity conditions, the power of our procedure is asymptotically one as sample size goes to infinity.

\subsection{FDR Control}

\begin{condition} \label{Condition1}
	\indent  The censoring time $U$ is independent of the irrelevant covariates in $\mathcal{S}_0^c$.
\end{condition}

Condition \ref{Condition1} is a mild condition and indicates that 
the censoring time can depend only on the relevant covariates in $\mathcal{S}_0$.  
The same condition was also used by \cite{dong2022reproducible} for accelerated failure time models.
In particular, this condition is automatically satisfied when the censoring time is independent of all the covariates.

\begin{theorem}\label{Theorem1}
	Under Condition~\ref{Condition1}, for any target FDR level $q\in(0, 1)$, our method with the knockoff threshold $\tau$ selecting the variables $\widehat{\mathcal{S}}_0=\left\{1\leq j\leq p: W_{j} \ge \tau\right\}$ 	
	controls the mFDR, that is,	$\mathrm{mFDR} = \mathbb{E} \left[|\widehat{\mathcal{S}}_0 \cap \mathcal{S}_0^c|/(|\widehat{\mathcal{S}}_0|+q^{-1})\right]\leq q$.
	Similarly, under Condition \ref{Condition1}, our method with the knockoff threshold $\tau_{+}$ selecting the variables $\widehat{\mathcal{S}}_0=\left\{1\leq j\leq p: W_{j} \ge \tau_{+}\right\}$ 	
	controls the exact FDR, that is, $\mathrm{FDR} = \mathbb{E} \left[|\widehat{\mathcal{S}}_0 \cap \mathcal{S}_0^c|/(|\widehat{\mathcal{S}}_0|\vee 1)\right]\leq q$.	
\end{theorem}

Theorem~\ref{Theorem1} means that both the modified and exact FDR controls are guaranteed for our method with the knockoff and knockoff+ thresholds, respectively. These results are non-asymptotic and hold for  
any sample size $n$ and any number of covariates $p$, meaning that our proposed method
can handle both the low-dimensional and high-dimensional regimes. 
None of the existing feature selection approaches for the Cox model can control the FDR unless the sample size tends to infinity.

It is easy to see from \eqref{eq: hat-b} and \eqref{eq: Statistic-Wj} that the LCD statistics $W_j$'s obey the flip-sign property for any regularization parameter $\lambda>0$, which says that swapping the $j$th covariate with its knockoff copy has the effect of changing the sign of $W_j$.  
Therefore, we can show that
the statistics corresponding to the irrelevant covariates have symmetric distributions about zero (see the proof of Theorem~\ref{Theorem1} later for details). It implies that for irrelevant covariates, the number
of statistics larger than any fixed threshold $t>0$ will be the same as that of statistics smaller than $-t$ in a probabilistic sense. This is the key to establishing the theorem above to guarantee the FDR control of our method.

\subsection{Power Analysis}

As mentioned earlier, there is no formal power analysis of the knockoffs framework for survival data in the literature. We aim to fill in this gap as a first attempt and provide theoretical foundations on the power
analysis for the knockoffs procedure in the Cox model.
Recall that $	\widetilde{\ell}(\bb)=n^{-1}\sum\limits_{i=1}^{n}\delta_i
\Big\{\bb^{\top}\bz_i-\log\Big[\sum\limits_{j=1}^{n}I(Y_j\geq Y_i) \exp(\bb^{\top}\bz_j)\Big] \Big\}$  with
$\bz_i=(\bx_i^{\top}, \widetilde{\bx}_i^{\top})^{\top}\in\mathbb{R}^{2p}$ for each $i=1. \ldots, n$. To ease the technical presentation, 
define
$\widetilde{\omega}_j(t, \bb)=I(Y_j\geq t)\exp(\bb^{\top}\bz_j)\,\,\,
\mbox{and}\,\,\,
\omega_j(t, \bb)=\widetilde{\omega}_j(t, \bb)/\Big[\sum\limits_{j=1}^n\widetilde{\omega}_j(t, \bb)\Big]$
for $j=1, \ldots, n$. 
Then, the Hessian matrix of 
$\widetilde{\ell}(\bb)$ is
\begin{align*}
\ddot{\widetilde{\ell}}(\bb) 
\equiv \frac{\partial^{2} \widetilde{\ell}(\bb)}{\partial \bb\partial \bb^{\top}}
=-n^{-1}\sum_{i=1}^n\delta_i
\Big\{\sum_{j=1}^n\omega_j(Y_i, \bb)\Big[\bz_j-\bar{\bz}(Y_i, \bb)\Big]^{\otimes 2}\Big\}, 
\end{align*}
where $\bar{\bz}(t, \bb)=\sum\limits_{j=1}^n\omega_j(t, \bb)\bz_j$ and $\balpha^{\otimes 2}=\balpha\balpha^{\top}$ for any column vector $\balpha$. 
Similar to \cite{fan2020rank}, in order to simplify the technical analysis, we assume that with asymptotic probability one, there are no ties in the magnitude of nonzero $W_j$'s and no ties in the magnitude of nonzero components of Lasso solution in \eqref{eq: hat-b}. To facilitate the power analysis, we impose the following regularity conditions.

\begin{condition}\label{Condition2}
	${\rm min}_{j \in \mathcal{S}_{0}}|\beta_{0j}| \geq \kappa_{n}n^{-1/2}\log^{1/2} (np)$ for some
	slowly diverging sequence $\kappa_n \to \infty$ as the sample size $n \to \infty$.
\end{condition}

\begin{condition} \label{Condition3}
	There exists some constant $c \in (2(qs_{0})^{-1},1)$ such that $|\widehat{\mathcal{S}}_0| \geq cs_{0}$ for $\widehat{\mathcal{S}}_0$ given in \eqref{eq: S0_hat} with asymptotic probability one, where $s_0$ is the number of nonzero elements of $\bbeta_0$ (i.e., $s_0=|\mathcal{S}_0|$). 
\end{condition}

\begin{condition} \label{Condition4}
	There exists some constant $K>0$ such that $P(|X_j|\leq K)=1$ for all $1\leq j\leq p$. 
\end{condition}

\begin{condition} \label{Condition5}
	Let $\bb_0=(\bbeta_0^{\top}, \0_p^{\top})^{\top}$ be a ($2p$)-vector with the last $p$ components being $0$ and
	\begin{align*}
	\kappa: =\kappa(s_0, 2)=\inf_{\left\{\v \in \mathbb{R}^{2p}\backslash \{0\}: \|\v_{\mathcal{S}_{0}^{c}}\|_{1} \leq 2 \|\v_{\mathcal{S}_{0}}\|_{1} \right\} } \frac{s_{0}^{1 / 2}\{\v^{\top} 	\ddot{\widetilde{\ell}}(\bb_0)\v\}^{1 / 2}}{\|\v_{\mathcal{S}_{0}}\|_{1}},  
	\end{align*} 
	where $\v_{\mathcal{S}_{0}}$ and $\v_{\mathcal{S}_{0}^{c}}$ are the subvectors 
	of $\v$ formed by components in sets $\mathcal{S}_{0}$ and $\mathcal{S}_{0}^{c}$, respectively. We have $1/\kappa=O_{p}(1)$.
\end{condition}

Condition~\ref{Condition2} puts a constraint on the minimum signal strength, which ensures that the Lasso solution $\widehat{\bb}(\lambda)$ given in \eqref{eq: hat-b} does not miss a great portion of important covariates in $\mathcal{S}_0$. Similar condition is widely used in the variable selection literature. In particular, if $p=p_n\to \infty$ as $n\to\infty$, this condition can be relaxed to ${\rm min}_{j \in \mathcal{S}_{0}}|\beta_{0j}| \geq \kappa_{n}\{(\log p)/n\}^{1/2}$, which is the same as Condition 2 in \cite{fan2020rank}.

Condition~\ref{Condition3} imposes a lower bound on the number of covariates selected by our knockoffs procedure. This is reasonable since any method with
high power should at least be able to select a large number of
covariates although these selected ones are not necessarily true relevant covariates. 
Conditions~\ref{Condition3} is also used in
\cite{fan2020rank} for power analysis in linear regression models.

Conditions~\ref{Condition4} assumes that all covariates are uniformly bounded, which holds in many real applications and 
was also widely used in the survival analysis literature, see, for example, \cite{bradic2011regularization}, \cite{fang2017testing}, \cite{yu2021confidence},  
and references therein. This assumption
is imposed to simplify our technical presentation and 
can be weakened to a  
tail condition on $X_j$ at the cost of much more tedious technical arguments.

Conditions~\ref{Condition5}, known as the compatibility factor condition~\citep{van2009conditions},  
is used to establish the upper bound of the $L_1$ estimation error of our Lasso solution $\widehat{\bb}(\lambda)$ given in \eqref{eq: hat-b}.  
It is common to put some assumption on the restricted eigenvalue (RE) or the compatibility factor to obtain oracle results for the Lasso. Here we adopt the same assumption in \cite{yu2021confidence} to simplify our presentation. Similar conditions also appear in
\cite{huang2013oracle} and \cite{fang2017testing}.

\begin{theorem}\label{Theorem2}
	Assume that  Conditions~\ref{Condition2}-\ref{Condition5} hold, $\lambda=C_{\lambda}\log^{1/2}(np) /n^{1/2}$ with $C_{\lambda}>0$ some constant, and $s_{0}={o}\{n^{1/2}/\log^{1/2} (n p)\}$. Then our knockoffs procedure satisfies that with
	asymptotic probability one,
	\begin{align}\label{eq: TDR-bound}
	|\widehat{\mathcal{S}}_0 \cap \mathcal{S}_0|/|\mathcal{S}_0|
	\geq 
	1-C_{\ell_1}C_{\lambda}(\varphi+1)\kappa_n^{-1}
	\end{align}
	and therefore
	\begin{align}\label{eq: power-bound}
	\mathrm{power}(\widehat{\mathcal{S}}_0 )=\mathbb{E}\left[|\widehat{\mathcal{S}}_0 \cap \mathcal{S}_0|/|\mathcal{S}_0|\right]
	\geq 
	1-C_{\ell_1}C_{\lambda}(\varphi+1)\kappa_n^{-1}
	\to 1
	\end{align}
	as $n\to\infty$, where $\varphi>0$ is the golden ratio satisfying $\varphi^2-\varphi-1=0$ and $C_{\ell_1}$ is some positive constant.  
\end{theorem}            

Theorem~\ref{Theorem2} reveals that our knockoffs-based feature selection procedure, CoxKnockoff, can have asymptotic power one under some mild regularity conditions. 
In view of \eqref{eq: power-bound}, we can conclude that the
stronger the signal, the faster the convergence of power to one because parameter $\kappa_n$ characterizes the signal strength.  In addition, as mentioned earlier, if $p=p_n\to\infty$ as $n\to\infty$, Condition~\ref{Condition2} can be relaxed to ${\rm min}_{j \in \mathcal{S}_{0}}|\beta_{0j}| \geq \kappa_{n}\{(\log p)/n\}^{1/2}$. Then in this case, it can be seen from the proof of Theorem~\ref{Theorem2} that the results in \eqref{eq: TDR-bound} and \eqref{eq: power-bound} still hold for $\lambda=C_{\lambda}\{(\log p)/n\}^{1/2}$ with some large constant $C_{\lambda}>0$. 
Combining the results in Theorems~\ref{Theorem1} and~\ref{Theorem2} shows that 
CoxKnockoff can enjoy appealing FDR control and power properties simultaneously.

\section{Simulation studies}\label{sec: simulations}

In this section, we conduct two simulation studies to verify our theoretical results
and examine the
finite-sample feature selection performance of CoxKonckoff in terms of both FDR and power. We will compare CoxKonckoff with the method in \cite{tibshirani1997lasso}, denoted by  CoxLasso, which is also based on the penalized partial log-likelihood with the $L_1$ penalty, but uses the original covariates only. 
To differentiate, we refer to our procedure
using threshold $\tau$ in \eqref{eq: tau} and threshold $\tau_{+}$ in \eqref{eq: tau+} as CoxKnockoffs and CoxKnockoffs+, respectively.
We 
evaluate the performance under different settings by varying signal strength, the distribution of covariates, the sample size, the number of covariates, and the censoring rate.
We use the equicorrelated construction to choose the vector $\bs$ in \eqref{eq: cov} when creating knockoff variables.  
The regularization parameter $\lambda$ is tuned by ten-fold cross-validation in all simulation studies.
Following many existing works on FDR control, such as~\cite{barber2015controlling},  \cite{fan2020rank}, and~\cite{fan2020ipad}, we set the target FDR level at $q=0.2$ for
	all simulation studies, meaning that we hope to control the FDR at a desired level of $0.2$. More numerical studies are provided in Appendix S2 of the Supporting Information to save space.

{\bf Study 1}. The first simulation study aims to examine the low-dimensional case where $p$ is comparable to but smaller than $n$. We set $(n, p)=(1000, 100)$ and consider three cases for the true coefficient vector $\bbeta_0=(\beta_{01}, \ldots, \beta_{0p})^{\top}$:
1) Case 1, $\beta_{0j}=5$ for $1\leq j\leq 10$ and $0$; 2) Case 2, $\beta_{0j}=2$ for $1\leq j\leq 10$ and $0$ otherwise; 3) Case 3, $\beta_{0j}=2$ for $1\leq j\leq 20$ and $0$ otherwise.
We also consider Gaussian distribution and multivariate $t$-distribution for the underlying distribution of the covariate vector $\bx$. For the former, each covariate vector $\bx_i$ is independently generated from a Gaussian distribution $N(\mathbf{0},\bSigma)$ for $i=1, \ldots, n$, where $\bSigma=(\Sigma_{kl})$ with $\Sigma_{kl}=0.5^{|k-l|}$ for $1\leq k, \ell\leq p$. For the latter, each $\bx_i$ is independently generated from the $t$-distribution with mean zero and degrees of freedom $\nu=3$. To be more specific,  
$\bx_i\thicksim [(\nu-2)/\nu]^{1/2}\u_i/\sqrt{\Gamma_i}$ for $i=1, \ldots, n$,
where $\u_i\sim N(\textbf{0}, \bSigma)$ with $\bSigma$ given before,
$\Gamma_i$ is independently sampled from a gamma distribution with the shape and rate parameter both equal to $\nu/2$. 
Given the covariate vector $\bx_i$, the survival time 
$T_i$ is generated based on the hazard function
$h(t|\bx_i) = h_0(t)\exp(\bbeta_0^{\top}\bx_i)$ with $h_0(t)=0.5$,
under which the survival time $T_i$ follows an exponential distribution with the rate parameter $0.5\exp(\bbeta_0^{\top}\bx_i)$.
The censoring times $U_i$'s are independently generated from an exponential distribution with parameter $\lambda_{c}$, which is chosen to obtain a censoring rate of about $20\%$ and $40\%$, respectively.

\begin{center}
	\begin{table*}[t]%
		\caption{The empirical FDR and power of different methods over 100 replications for Study 1 
			and target FDR level $q=0.2$. \label{tab-n1000-p100}}
		\centering
		\begin{tabular}{lll ll ll ll}
			\toprule	
			\multirow{2}{*}{\textbf{Distribution}} & \multirow{2}{*}{\textbf{CR}} & \multirow{2}{*}{\textbf{Method}} & \multicolumn{2}{c}{\textbf{Case 1}} & \multicolumn{2}{c}{\textbf{Case 2}} & \multicolumn{2}{c}{\textbf{Case 3}}  \\ [-9pt]
			& & & \multicolumn{2}{c@{}}{\hrulefill} & \multicolumn{2}{c@{}}{\hrulefill} & \multicolumn{2}{c@{}}{\hrulefill} \\
			& & &  \textbf{FDR} & \textbf{Power}   & \textbf{FDR} & \textbf{Power}  & \textbf{FDR} & \textbf{Power} \\
			\hline
			Gaussian & 20$\%$  & CoxLasso     & 0.878  & 1   & 0.832   & 1   & 0.748   & 1  \\
			&         & CoxKnockoff  & 0.237  & 1   & 0.247   & 1   & 0.219   & 1    \\
			&         & CoxKnockoff+ & 0.183  & 1   & 0.179   & 1   & 0.177   & 1  \\
			& 40$\%$  & CoxLasso     & 0.872  & 1   & 0.824   & 1   & 0.740   & 1 \\
			&         & CoxKnockoff  & 0.217  & 1   & 0.252   & 1   & 0.235   & 1 \\
			&         & CoxKnockoff+ & 0.162  & 1   & 0.190   & 1   & 0.197   & 1  \\
			\midrule
			$t$  & 20$\%$  & CoxLasso       & 0.660  & 1       & 0.684   & 1  & 0.579   & 1  \\
			&         & CoxKnockoff    & 0.211  & 1       & 0.197   & 1  & 0.185   & 1 \\
			&         & CoxKnockoff+   & 0.157  & 1       & 0.151   & 1  & 0.152   & 1 \\
			& 40$\%$  & CoxLasso       & 0.615  & 0.999   & 0.655   & 1  & 0.539   & 0.999 \\
			&         & CoxKnockoff    & 0.163  & 0.999   & 0.192   & 1  & 0.165   & 0.998 \\
			&         & CoxKnockoff+   & 0.125  & 0.999   & 0.137   & 1  & 0.145   & 0.998 \\
			\bottomrule
		\end{tabular}
	\end{table*}
\end{center}

Table~\ref{tab-n1000-p100} summarizes the empirical FDR and power based on 100 replications for different methods 
under various scenarios in Study 1. 
It can be seen that all methods enjoy high power, meaning that all methods can discover the truly relevant features. However, in terms of FDR, our methods (CoxKnockoff and CoxKnockoff+) are obviously better than CoxLasso in all settings. To be more specific, CoxKnockoff+  
can strictly control the FDR below the target level. Some FDR results of CoxKnockoff 
are slightly higher than the target level. This is reasonable since the knockoff threshold is designed for controlling the mFDR. By contrast, the CoxLasso method fails to control the FDR 
since all the FDR values are greater than 0.5, which means that more than $50\%$ of the identified features by CoxLasso are irrelevant.

{\bf Study 2}. In the second simulation study, we further investigate the finite-sample performance under the high-dimensional case 
where $p$ is larger than $n$.
We use the same model setup as before, except that $(n, p)=(400, 600), (400, 800)$ and $(400, 1000)$. 
Table~\ref{tab-Study2} reports the empirical FDR and power based on 100 replications for different methods 
under various scenarios in Study 2. 
It is seen that CoxKnockoff+ again achieves the best performance in terms of FDR and power in all settings. The performance of CoxKnockoff+ is robust
across different settings.
CoxKnockoff also has a very high power and sometimes obtains a FDR value slightly higher than the target level.
By contrast, CoxLasso fails to control the FDR again although it has a very high power. Its FDR is far above the nominal level. In fact, all the FDR values of CoxLasso are greater than 0.7.

\begin{center}
	\begin{table*}[h!]%
		\caption{The empirical FDR and power of different methods over 100 replications for Study 2 and target FDR level $q=0.2$.}
		\label{tab-Study2}
		\centering
		\scalebox{0.86}{
		\begin{tabular}{llll cl cl cl}
			\toprule	
			\multirow{2}{*}{\textbf{Distribution}} & \multirow{2}{*}{$(n,p)$} & 
			\multirow{2}{*}{\textbf{CR}} & \multirow{2}{*}{\textbf{Method}} & \multicolumn{2}{c}{\textbf{Case 1}} & \multicolumn{2}{c}{\textbf{Case 2}} & \multicolumn{2}{c}{\textbf{Case 3}}  \\ [-9pt] 
			& & & & \multicolumn{2}{c@{}}{\hrulefill} & \multicolumn{2}{c@{}}{\hrulefill} & \multicolumn{2}{c@{}}{\hrulefill}\\
			&  &  &  & \textbf{FDR} & \textbf{Power}   & \textbf{FDR} & \textbf{Power}  & \textbf{FDR} & \textbf{Power}  \\ 
			\hline
			Gaussian	& $(400,600)$ & 20$\%$  & CoxLasso     & 0.932  & 1  & 0.889  & 1  & 0.853   & 1  \\
			&             &         & CoxKnockoff  & 0.208  & 1  & 0.233  & 1  & 0.214   & 1  \\
			&             &         & CoxKnockoff+ & 0.146  & 1  & 0.173  & 1  & 0.180   & 1  \\
			&             & 40$\%$  & CoxLasso     & 0.923  & 1  & 0.882  & 1  & 0.834   & 1 \\
			&             &         & CoxKnockoff  & 0.211  & 1  & 0.201  & 1  & 0.210   & 1 \\
			&             &         & CoxKnockoff+ & 0.157  & 1  & 0.152  & 1  & 0.164   & 1  \\
			\midrule
			Gaussian	& $(400,800)$ & 20$\%$  & CoxLasso     & 0.938  & 1  & 0.902  & 1  & 0.863   & 1  \\
			&             &         & CoxKnockoff  & 0.231  & 1  & 0.206  & 1  & 0.221   & 1  \\
			&             &         & CoxKnockoff+ & 0.175  & 1  & 0.151  & 1  & 0.190   & 1  \\
			&             & 40$\%$  & CoxLasso     & 0.930  & 1  & 0.890  & 1  & 0.849   & 1 \\
			&             &         & CoxKnockoff  & 0.241  & 1  & 0.233  & 1  & 0.225   & 1 \\
			&             &         & CoxKnockoff+ & 0.186  & 1  & 0.163  & 1  & 0.183   & 1  \\
			\midrule
			Gaussian	&$(400,1000)$ & 20$\%$ & CoxLasso     & 0.942  & 1  & 0.906  & 1  & 0.871   & 1  \\
			&             &        & CoxKnockoff  & 0.256  & 1  & 0.251  & 1  & 0.214   & 1  \\
			&             &        & CoxKnockoff+ & 0.195  & 1  & 0.194  & 1  & 0.189   & 1  \\
			&             & 40$\%$ & CoxLasso     & 0.933  & 1  & 0.898  & 1  & 0.857   & 1 \\
			&             &        & CoxKnockoff  & 0.239  & 1  & 0.229  & 1  & 0.180   & 1 \\
			&             &        & CoxKnockoff+ & 0.181  & 1  & 0.171  & 1  & 0.155   & 1  \\			
			\midrule
			$t$  & $(400,600)$ & 20$\%$  & CoxLasso    & 0.845  & 1    & 0.783   & 1      & 0.727   & 0.995 \\
			&             &         & CoxKnocoff  & 0.136  & 1    & 0.104   & 1      & 0.136   & 0.994 \\
			&             &         & CoxKnocoff+ & 0.087  & 1    & 0.072   & 1      & 0.114   & 0.994 \\
			&             & 40$\%$  & CoxLasso    & 0.805  & 1    & 0.779   & 1      & 0.715   & 1 \\
			&             &         & CoxKnocoff  & 0.132  & 1    & 0.134   & 1      & 0.118   & 1 \\
			&             &         & CoxKnocoff+ & 0.097  & 1    & 0.092   & 1      & 0.095   & 1 \\ 	 
			\midrule
			$t$   & $(400,800)$ & 20$\%$  & CoxLasso    & 0.770  & 1    & 0.804  & 1  & 0.730   & 1  \\
			&             &         & CoxKnocoff  & 0.139  & 1    & 0.123  & 1  & 0.104   & 1 \\
			&             &         & CoxKnocoff+ & 0.104  & 1    & 0.099  & 1  & 0.090   & 1 \\
			&             & 40$\%$  & CoxLasso    & 0.785  & 1    & 0.770  & 1  & 0.725   & 0.999 \\
			&             &         & CoxKnocoff  & 0.134  & 1    & 0.110  & 1  & 0.115   & 0.999 \\
			&             &         & CoxKnocoff+ & 0.103  & 1    & 0.080  & 1  & 0.099   & 0.999 \\
			\midrule
			$t$   & $(400,1000)$ & 20$\%$ & CoxLasso    & 0.789  & 1   & 0.804   & 0.999  & 0.762   & 1  \\
			&             &         & CoxKnocoff  & 0.151  & 1   & 0.128   & 0.999  & 0.121   & 1 \\
			&             &         & CoxKnocoff+ & 0.118  & 1   & 0.096   & 0.999  & 0.107   & 1 \\
			&             & 40$\%$  & CoxLasso    & 0.831  & 1   & 0.812   & 1      & 0.714   & 1 \\
			&             &         & CoxKnocoff  & 0.121  & 1   & 0.114   & 1      & 0.107   & 0.999 \\
			&             &         & CoxKnocoff+ & 0.099  & 1   & 0.083   & 1      & 0.093   & 0.999 \\ 			
			\bottomrule
		\end{tabular}}
	\end{table*}
\end{center}

\section{Real data analysis}\label{sec: RealDataAnalysis}

We finally evaluate the performance of the proposed method on a breast cancer data set 
analyzed in~\cite{van2002gene}. This data set, named~\texttt{nki70} in the R package~\texttt{penalized}, 
includes 144 lymph node positive breast cancer patients.
The observed time is the variable ``time'' in the data set, which denotes Metastasis-free follow-up time in months. The variable ``event'' records 
the status of the patient at the end of study and patients with value of $0$ for the variable ``event'' are considered censored.
The censoring rate is around $67\%$. 
The covariates used in our Cox model consists of five risk factors, Diam (diameter of tumor), N (number of affected lymph nodes), ER (Estrogen receptor status), Grade (Grade of the tumor), and Age (Patient age at diagnosis), 
and 70 measures of gene expression, resulting in a sample of $n=144$ observations with $p=75$ covariates. 
To identify important covariates associated with Metastasis-free follow-up time, we employ three approaches (CoxLasso, CoxKnockoff, and CoxKnockoff+) with target FDR level $q = 0.2$.
The regularization parameter $\lambda$ 
is chosen by five-fold cross-validation.

\begin{table*}[h!]
	\centering
	\caption{Feature selection results for the Breast cancer data.}
	\begin{tabular}{l l} 
		\toprule
		\textbf{Method} & \textbf{Selected covariates} \\ 
		\hline
		CoxLasso     & N, Age, QSCN6L1, Contig32125$\_$RC, ZNF533, IGFBP5.1, PRC1\\
		CoxKnockoff  & N, QSCN6L1, ZNF533, IGFBP5.1, PRC1\\
		CoxKnockoff+ & N, QSCN6L1, ZNF533, IGFBP5.1, PRC1\\
		\bottomrule
	\end{tabular}\label{tab-nki70}
\end{table*}

Table~\ref{tab-nki70} lists the selected covariates by different methods. It is seen that CoxKnockoff and CoxKnockoff+ both selected five covariates,  all of which were also selected by CoxLasso. The CoxLasso method identified two more covariates, Age and Contig32125$\_$RC.
However, \cite{mook200970} and \cite{liu2012variable} found that age was not significant when they analyzed other lymph node positive breast cancer data sets.

\section{Conclusions}\label{sec: Discussion}

In this paper, we have proposed a new and flexible feature selection framework for the Cox model with theoretical guarantees in FDR control and power. We have established that the proposed method can control the FDR in finite sample with any dimension and the power is asymptotically one as sample size tends to infinity.  Compared with the existing feature selection methods for the Cox model, our procedure is guaranteed to control the FDR for any sample size. In addition, our method has expanded the scope of the currently fast growing area of knockoffs by providing the formal theoretical justification on the power behavior in the survival setting. This is also new to the literature.

We have focused on the Cox model in this paper. Our idea of using  
knockoffs for controlled feature selection can be extended to other models for censored data, such as
partially linear Cox model~\citep{wu2020variable}  
and high-dimensional additive hazards models~\citep{lin2013high, zheng2022L0}.
These possible extensions are beyond the scope of the current paper and will be interesting topics for future research.



\newpage
\setcounter{page}{1}



\begin{center}{\bf \Large Supporting Information to ``CoxKnockoff: Controlled Feature Selection for the Cox Model Using Knockoffs''}

	\bigskip
	
	Daoji Li$^1$,  Jinzhao Yu$^2$, and Hui Zhao$^2$
	
	$^1$Department of Information Systems and Decision Sciences, California State University, Fullerton, CA, 92831, United States
	
	$^2$School of Statistics and Mathematics, Zhongnan University of Economics and Law, Wuhan, 430073, China\\

\end{center}


\setcounter{equation}{0}
\setcounter{table}{0}
\renewcommand{\theequation}{A.\arabic{equation}}
\renewcommand{\thetable}{S\arabic{table}}


This Supporting Information consists of two parts. The first part includes all the proofs of the main results and technical details while the second part provides additional numerical studies.

	\section*{Appendix S1: Proofs of the main results and technical details}\label{App: S1}

We start with some notation. Let $\widetilde{\X}=(\widetilde{\bx}_1, \ldots, \widetilde{\bx}_n)^{\top}$ and $\X=(\bx_1, \ldots, \bx_n)^{\top}$, where $\widetilde{\bx}_i$ is the model-X knockoff copy of $\bx_i$ for $i=1, \ldots, n$. Write $\y=(Y_1, \ldots, Y_n)^{\top}$ and
$\bdelta=(\delta_1, \ldots, \delta_n)^{\top}$.	
Before proving our Theorems~\ref{Theorem1} and~\ref{Theorem2}, we first present two Lemmas, which are extracted from 
Lemma 1 of \cite{dong2022reproducible} and Lemma 1(ii) of \cite{yu2021confidence}, respectively. 
Lemma~\ref{lemma: Exchangeability} shows that swapping any irrelevant covariates with their knockoff copies will not change the conditional joint distribution of the original covariates and their knockoffs given both the censored response variable and the censoring indicator.  Lemma~\ref{lemma: hatb-estimation-error} provides the  upper bound of the $L_1$ estimation error of our Lasso solution $\widehat{\bb}(\lambda)$ given in \eqref{eq: hat-b}.  
Lemma~\ref{lemma: Exchangeability} is needed 
in the proof of our Theorem~\ref{Theorem1} while  Lemma~\ref{lemma: hatb-estimation-error} will be used in the proof of our Theorem~\ref{Theorem2}.

\begin{lemma}\label{lemma: Exchangeability}
	\emph{(Lemma 1 of \cite{dong2022reproducible})} Under Condition~\ref{Condition1}, we have  $[\mathbf{X},\widetilde{\mathbf{X}}]|(\y^{\top},\bdelta^{\top})\overset{d}=
	[\mathbf{X},\widetilde{\mathbf{X}}]_{\mathrm{swap}(\mathcal{S})}|(\y^{\top},\bdelta^{\top})$ for any set $\mathcal{S} \subset \mathcal{S}_{0}^c$, where $[\mathbf{X},\widetilde{\mathbf{X}}]_{\mathrm{swap}(\mathcal{S})}$ is the matrix obtained by swapping columns in $\mathcal{S}$.
\end{lemma}

\begin{lemma}\label{lemma: hatb-estimation-error}	
	\emph{(Lemma 1(ii) of \cite{yu2021confidence})} Assume that Conditions~\ref{Condition4} and \ref{Condition5} hold. If $s_0=o\{n^{1/2}/\log^{1/2}(np)\}$ and $\lambda=C_{\lambda}n^{-1/2}\log^{1/2}(np)$ with $C_{\lambda}$ some positive constant, then we have
	\begin{align*}
	&     \|\widehat{\bb}(\lambda)-\bb_0\|_1
	\leq C_{\ell_1}\lambda s_0.
	\end{align*}
\end{lemma}

\noindent{\bf Remark 1}.  Note that Lemma~\ref{lemma: hatb-estimation-error} holds even when $p$ is finite. As pointed out by \cite{yu2021confidence}, if $p=p_n\to\infty$ as $n\to\infty$, the result in this lemma still holds for  $\lambda=C_{\lambda}\{(\log p)/n\}^{1/2}$ with some large constant $C_{\lambda}>0$.

\vspace{8mm}
\noindent{\bf Proof of Theorem~\ref{Theorem1}.}  We first claim that conditional on $\{|W_1|, \cdots, |W_p|\}$,  the signs of $W_j$'s, $j\in \mathcal{S}_0^c$, are independent and identically distributed coin flips. In view of \eqref{eq: hat-b} and \eqref{eq: Statistic-Wj}, we can see that the statistic $W_j$ depends on the original covariates, the knockoffs, the response, and the censoring indicator. In other words,  each statistic $W_j$ can be rewritten as
\begin{align*}
W_j = \mathrm{w}_j([\mathbf{X},\widetilde{\mathbf{X}}],\y,\bdelta)
\end{align*}
with some function $\mathrm{w}_j$. 

Let  $\epsilon=(\epsilon_{1},\ldots,\epsilon_{p})$
be a sequence of independent random variables
such that $\epsilon_{j}=1$ or $-1$ with equal probability of $0.5$ for $j\in\mathcal{S}_0^c$ and $1$ otherwise. Write $W=(W_1, \ldots, W_p)$. Then, to prove 
the claim above, it is sufficient to show that
\begin{align}\label{eq: W-flip-coin}
W \overset{d}= \epsilon \odot W,
\end{align}
where $\odot$ denotes the pointwise multiplication, that is, $\epsilon \odot W=(\epsilon_{1} W_{1}, \ldots, \epsilon_{p} W_{p})$. 

Define the subset $\mathcal{S}=\left\{j: \epsilon_{j}=-1\right\}$. By the definition of $\epsilon_{j}$, it holds that $\mathcal{S} \subset \mathcal{S}_{0}^c$. Then we swap variables $X_j$'s in $\mathcal{S}$ with their knockoffs and let
\begin{align*}
W_{\mathrm{swap}(\mathcal{S})}\triangleq W([\mathbf{X}, \widetilde{\mathbf{X}}]_{\mathrm{swap}(S)}, \y,\bdelta),
\end{align*}
where $[\mathbf{X},\widetilde{\mathbf{X}}]_{\mathrm{swap}(\mathcal{S})}$ is the matrix obtained by swapping columns in $\mathcal{S}$.

On the one hand, it follows from \eqref{eq: hat-b} and \eqref{eq: Statistic-Wj} that swapping the $j$th covariate with its knockoff copy only leads to the sign change of the $j$th statistic $W_j$ for any regularization parameter $\lambda>0$. In other words, our statistics $W_j$'s obey the flip-sign property. Therefore, we have $ W_{\mathrm{swap}(\mathcal{S})}=\epsilon \odot W$. On the other hand, under Condition~\ref{Condition1},  it follows from Lemma~\ref{lemma: Exchangeability} that $W_{\mathrm{swap}(\mathcal{S})} \overset{d}= W$ since $\mathcal{S}\subset \mathcal{S}_{0}^c$. Thus, \eqref{eq: W-flip-coin} holds and our claim is true. That is, conditional on $\{|W_1|, \cdots, |W_p|\}$,  the signs of $W_j$'s, $j\in \mathcal{S}_0^c$, are independent and identically distributed coin flips. Therefore, the statistics $W_j$'s corresponding to the irrelevant covariates have symmetric distributions across zero. This means that, for irrelevant covariates, the number of statistics larger than a threshold $\tau$ (or $\tau_{+}$) will be the same as that of statistics smaller than $-\tau$ (or $-\tau_{+}$) in a probabilistic sense. With this result, the remaining part of the proof of Theorem~\ref{Theorem1} in our paper is the same as that of Theorem~\ref{Theorem1} in \cite{candes2018panning}. Therefore, the details are omitted here to save space.

\vspace{8mm}
\noindent{\bf Proof of Theorem~\ref{Theorem2}.}
Recall that our procedure
using threshold $\tau$ in \eqref{eq: tau} and threshold $\tau_{+}$ in \eqref{eq: tau+} are denoted as CoxKnockoffs and CoxKnockoffs+, respectively.  Since the proofs for CoxKnockoffs and CoxKnockoffs+ are similar,  we only show the latter to save space. 

Under the assumptions  $s_0=o\{n^{1/2}/\log^{1/2}(np)\}$,  $\lambda=C_{\lambda}n^{-1/2}\log^{1/2}(np)$ with constant $C_{\lambda}>0$ some constant,  Condition~\ref{Condition4}, and Condition~\ref{Condition5}, it follows from Lemma~\ref{lemma: hatb-estimation-error} that 
\begin{align}\label{eq: oracle-inequality-beta} 
C_{\ell_1}\lambda s_0
\geq & \|\widehat{\bb}(\lambda)-\bb_0\|_1
=\sum_{j=1}^p\left[|\widehat{b}_j(\lambda)-\beta_{0j}|+ |\widehat{b}_{j+p}(\lambda)|\right],
\end{align}    
where $C_{\ell_1}$ is some positive constant 
and $\bb_0=(\bbeta_0^{\top}, \0_p^{\top})^{\top}$ is a ($2p$)-vector with the last $p$ components being 0. By Condition~\ref{Condition2} and the assumption  
$\lambda=C_{\lambda}n^{-1/2}\log^{1/2}(np)$,
we have
\begin{align}\label{eq: min-beta}
\min_{j\in\mathcal{S}_0}|\beta_{0j}|\geq \kappa_n\lambda/C_{\lambda}.
\end{align}

Recall that by assumption, there are no ties in the magnitude of nonzero $W_j$'s and no ties in the nonzero components of the Lasso solution in \eqref{eq: hat-b} with asymptotic probability one. Let $|W_{(1)}|\geq |W_{(2)}|\geq \cdots\geq  |W_{(p)}|$ be the ordered statistics of $W_1, \ldots, W_p$ according to magnitude. Denote by $j^{*}$ the index satisfying that $|W_{(j^{*})}|=\tau_{+}$. 
Then by the definition of $\tau_{+}$ in \eqref{eq: tau+}, we have $-\tau_{+}< W_{(j^{*}+1)}\leq 0$.
To show this, let us restate the definition of $\tau_{+}$ in \eqref{eq: tau+}, which is 
	\begin{align*}
	\tau_+ = \min\left\{t\in \mathcal{W}: \frac{|\{j: W_j \leq -t\}|+1}{|\{j: W_j \geq t\}|\vee 1}\leq q\right\},
	\end{align*}
	where $\mathcal{W}=\{|W_j|: 1\leq j\leq p\}\setminus\{0\}$. We
	first note that $\tau_{+}>0$ by the definition of $\tau_{+}$ and $|W_{(j^{*}+1)}|<|W_{(j^{*})}|=\tau_{+}$ because  $|W_{(j^{*}+1)}|=|W_{(j^{*})}|$ 
	would contradict with the assumption that there are no ties in the magnitude of nonzero $W_j$'s. Therefore, to prove 
	$-\tau_{+}< W_{(j^{*}+1)}\leq 0$, it is sufficient to show that $ 0<W_{(j^{*}+1)}<\tau_{+}$ cannot hold by contradiction.  Suppose $0<W_{(j^{*}+1)}<\tau_{+}$. 
	Then, by the definition of $\tau_{+}$, we have
	\begin{align}\label{eq: greater-than-q}
	\frac{|\{j: W_j \leq -W_{(j^{*}+1)}\}|+1}{|\{j: W_j \geq W_{(j^{*}+1)}\}|\vee 1}> q.
	\end{align}
	Next, we show that $\{j: -\tau_{+} <W_j \leq -W_{(j^{*}+1)}\}$ is an empty set. If this is not true, there exists some 
	\begin{align}\label{eq: j0}
	j_0\in\{j: -\tau_{+} <W_j \leq -W_{(j^{*}+1)}\}\subseteq \{1, \cdots, p\},
	\end{align} 
	which leads to $-\tau_{+} <W_{j_0} \leq -W_{(j^{*}+1)}$.  Recall that $|W_{(1)}|\geq |W_{(2)}|\geq \cdots\geq  |W_{(p)}|$ are the ordered statistics of $W_1, \ldots, W_p$ according to magnitude. Let $k_1, \cdots, k_p$ be a permutation of $\{1, \cdots, p\}$ such that $W_{k_j}=W_{(j)}$. 
	On one hand, by $W_{j_0} \leq -W_{(j^{*}+1)}$,  $W_{(j^{*}+1)}>0$, and $|W_{(1)}|\geq |W_{(2)}|\geq \cdots\geq  |W_{(p)}|$, we have $|W_{j_0}|\neq 0$ and $|W_{j_0}| \geq |W_{(j^{*}+1)}|\geq \cdots\geq |W_{(p)}|$. This entails that $j_0\notin\{k_{j*+1}, \cdots, k_p\}$. To see this, 
	assume that $j_0=k_{j*+\ell}$ with some $1\leq \ell\leq p-j^{*}$, then $|W_{j_0}|=|W_{k_{j^*+\ell}}|=|W_{(j^*+\ell)}|$, which contradicts with the assumption that there are no ties in the magnitude of nonzero $W_j$'s. On the other hand, by $-\tau_{+} <W_{j_0} \leq -W_{(j^{*}+1)}$, $W_{(j^{*}+1)}>0$ and $|W_{(j^{*})}|=\tau_{+}$, we have $|W_{j_0}|< |W_{(j^{*})}|$, implying that $j_0\notin\{k_{1}, \cdots, k_{j*}\}$. Combining the above results yields that $j_0\notin\{1, \cdots, k_{j*}\}\cup\{k_{j*+1}, \cdots, k_p\}=\{k_1, \cdots, k_p\}=\{1, \cdots, p\}$, which contradicts with \eqref{eq: j0}. Therefore, $\{j: -\tau_{+} <W_j \leq -W_{(j^{*}+1)}\}$ is an empty set under the assumption $0<W_{(j^*+1)}<\tau_{+}$ and then
	\begin{align}\label{eq: empty-set}
	\{j: W_j \leq -W_{(j^{*}+1)}\} 
	= \{j: W_j \leq -\tau_{+}\}\cup \{j: -\tau_{+} <W_j \leq -W_{(j^{*}+1)}\}
	=\{j: W_j \leq -\tau_{+}\}.
	\end{align}
	Moreover, by the assumption $0<W_{(j^*+1)}<\tau_{+}$, we have $\{j: W_j \geq \tau_{+}\}\subseteq \{j: W_j \geq W_{(j^{*}+1)}\}$. This, together with  \eqref{eq: empty-set}, entails   
	\begin{align}\label{eq: smaller-than-q}
	\frac{|\{j: W_j \leq -W_{(j^{*}+1)}\}|+1}{|\{j: W_j \geq W_{(j^{*}+1)}\}|\vee 1}
	= \frac{|\{j: W_j \leq -\tau_{+}\}|+1}{|\{j: W_j \geq W_{(j^{*}+1)}\}|\vee 1}
	\leq \frac{|\{j: W_j \leq -\tau_{+}\}|+1}{|\{j: W_j \geq \tau_{+}\}|\vee 1}\leq q,
	\end{align}
	where the last inequality follows from the definition of $\tau_{+}$.  Since  results \eqref{eq: greater-than-q} and \eqref{eq: smaller-than-q} contradict with each other, we can conclude that the assumption $0<W_{(j^{*}+1)}<\tau_{+}$ cannot hold.  Therefore, we have $-\tau_{+}< W_{(j^{*}+1)}\leq 0$.
In order to proceed, we next consider two cases $-\tau_{+}< W_{(j^{*}+1)}< 0$ and $W_{(j^{*}+1)}=0$  separately.

{\bf Case 1}: $-\tau_{+}< W_{(j^{*}+1)}< 0$. By the definition of $\tau_{+}$ in \eqref{eq: tau+} and $j^{*}$, we have 
\begin{align}\label{eq: Case2_eq1} 
\frac{|\{j: W_j \leq -\tau_{+}\}|+2}{|\{j: W_j \geq \tau_{+}\}|\vee 1}>q
\end{align}
since otherwise we can get a new threshold $|W_{(j^{*}+1)}|$, which is smaller than $\tau_{+}$. 
Thus, it follows from \eqref{eq: Case2_eq1} that 
\begin{align*} 
|\{j: W_j \leq -\tau_{+}\}|
\geq q(|\{j: W_j \geq \tau_{+}\}|\vee 1)-2
\geq q|\{j: W_j \geq \tau_{+}\}|-2
=q|\widehat{\mathcal{S}}_0|-2.
\end{align*}
Then by Condition~\ref{Condition3}, we have $|\widehat{\mathcal{S}}_0|\geq cs_0$ and thus 
\begin{align*} 
|\{j: W_j \leq -\tau_{+}\}|
\geq qcs_0-2
\end{align*}
with asymptotic probability one. Furthermore, by the definition of $W_j$ in \eqref{eq: Statistic-Wj}, when $W_j\leq -\tau_{+}$, we have $|\widehat{b}_j(\lambda)| - |\widehat{b}_{j+p}(\lambda)|\leq -\tau_{+}$ and thus $|\widehat{b}_{j+p}(\lambda)|\geq |\widehat{b}_j(\lambda)|+\tau_{+}\geq \tau_{+}$. This, together with \eqref{eq: oracle-inequality-beta}, entails that
\begin{align}\label{eq: A5}
C_{\ell_1}\lambda s_0
\geq & \sum_{j:\, W_j \leq -\tau_{+}}|\widehat{b}_{j+p}(\lambda)|
\geq \tau_{+}|\{j: W_j \leq -\tau_{+}\}|.
\end{align}
Combining these results yields $C_{\ell_1}\lambda s_0\geq \tau_{+}(qcs_0-2)$ and thus
\begin{align}\label{eq: tau+_bound}
\tau_{+}\leq 
C_{\ell_1}\lambda s_0/(qcs_0-2)\leq \kappa_n\lambda/(C_{\lambda}\varphi)
\end{align}
for large enough $n$ since $\kappa_n\to\infty$ as $n\to\infty$.  Here $\varphi$ is the golden ratio, which is the positive solution of $\varphi^2-\varphi-1=0$.

Recall that $\widehat{\mathcal{S}}_0=\{1\leq j\leq p: W_j\geq \tau_{+}\}$. Then, by the definition of $W_j$ in \eqref{eq: Statistic-Wj}, we have $|\widehat{b}_j(\lambda)| - |\widehat{b}_{j+p}(\lambda)|< \tau_{+}$ for any $j\in (\widehat{\mathcal{S}}_0)^c$. This, together with \eqref{eq: oracle-inequality-beta}, entails that
\begin{align*}
C_{\ell_1}\lambda s_0
\geq \sum_{j\in \mathcal{S}_0\cap (\widehat{\mathcal{S}}_0)^c}\left[|\widehat{b}_j(\lambda)-\beta_{0j}|+ |\widehat{b}_{j+p}(\lambda)|\right]
> \sum_{j\in \mathcal{S}_0\cap (\widehat{\mathcal{S}}_0)^c}\left[|\widehat{b}_j(\lambda)-\beta_{0j}|+ |\widehat{b}_{j}(\lambda)|-\tau_{+}\right].
\end{align*}
By the triangle inequality and \eqref{eq: min-beta}, we can conclude that
\begin{align*}
C_{\ell_1}\lambda s_0
> \sum_{j\in \mathcal{S}_0\cap (\widehat{\mathcal{S}}_0)^c}\left(|\beta_{0j}|-\tau_{+}\right)
\geq (C_{\lambda}^{-1}\lambda\kappa_n-\tau_{+})|\mathcal{S}_0\cap (\widehat{\mathcal{S}}_0)^c|.
\end{align*}
Using this result and \eqref{eq: tau+_bound}, we have
\begin{align}\label{eq: case1-bound}
\frac{|\widehat{\mathcal{S}}_0 \cap \mathcal{S}_0|}{|\mathcal{S}_0|}
=1- \frac{|\mathcal{S}_0\cap (\widehat{\mathcal{S}}_0)^c|}{s_0}
> 1- \frac{C_{\ell_1}\lambda}{C_{\lambda}^{-1}\lambda\kappa_n-\tau_{+}} 
\geq 1-\frac{C_{\ell_1}C_{\lambda}\varphi}{\kappa_n(\varphi-1)}.
\end{align}

{\bf Case 2}: $W_{(j^{*}+1)}=0$. Since $|W_{(1)}|\geq |W_{(2)}|\geq \cdots\geq  |W_{(p)}|$, we have $|W_{(j^{*}+1)}|= \cdots=|W_{(p)}|=0$. Thus the estimated set $\widehat{\mathcal{S}}_0=\{1\leq j\leq p: W_j\geq \tau_{+}\}=\{1\leq j\leq p: W_j>0\}$ and the set $\{1\leq j\leq p: W_j\leq -\tau_{+}\}=\{1\leq j\leq p: W_j<0\}$. 
Write $\mbox{supp}(W)=\{1\leq j\leq p: W_j\neq 0\}$ and  $\mathcal{S}_{1}=\{1\leq j\leq p: W_j< 0\}$. 
Then we have $\widehat{\mathcal{S}}_0=\mbox{supp}(W)\backslash\mathcal{S}_{1}$.

If $|\mathcal{S}_{1}|>\varphi C_{\ell_1}C_{\lambda}\kappa_n^{-1}s_0$, then 
we can show that $\tau_{+}\leq \kappa_n\lambda/(C_{\lambda}\varphi)$ for large enough $n$ using the same arguments as in \eqref{eq: A5} since $\{1\leq j\leq p: W_j\leq -\tau_{+}\}=\{1\leq j\leq p: W_j<0\}$. Thus, it reduces to Case 1 above
and the arguments therein follow.

If $|\mathcal{S}_1|\leq \varphi C_{\ell_1}C_{\lambda}\kappa_n^{-1}s_0$, then we have
\begin{align}\label{eq: bound-S0hat-S0}
|\widehat{\mathcal{S}}_0 \cap \mathcal{S}_0|
=|\mbox{supp}(W) \cap \mathcal{S}_0|-	|\mathcal{S}_1 \cap \mathcal{S}_0|
\geq |\mbox{supp}(W) \cap \mathcal{S}_0|-	\varphi C_{\ell_1}C_{\lambda}\kappa_n^{-1}s_0
\end{align}
by noting that $\widehat{\mathcal{S}}_0=\mbox{supp}(W)\backslash\mathcal{S}_{1}$.  Next, we will find the lower bound of $|\mbox{supp}(W) \cap \mathcal{S}_0|$. Define $\mathcal{S}_2=\{1\leq j\leq p: \widehat{b}_j(\lambda)=0\}$. 
It follows from \eqref{eq: oracle-inequality-beta} and \eqref{eq: min-beta} that
\begin{align*}
C_{\ell_1}\lambda s_0
\geq \sum_{j\in \mathcal{S}_2\cap\mathcal{S}_0}|\widehat{b}_j(\lambda)-\beta_{0j}|
=\sum_{j\in \mathcal{S}_2\cap\mathcal{S}_0}|\beta_{0j}|
\geq |\mathcal{S}_2\cap\mathcal{S}_0|\min_{j\in\mathcal{S}_0}|\beta_{0j}|
\geq |\mathcal{S}_2\cap\mathcal{S}_0|(\kappa_n\lambda/C_{\lambda})
\end{align*}
which leads to $	|\mathcal{S}_2\cap\mathcal{S}_0|
\leq  C_{\ell_1}C_{\lambda}\kappa_n^{-1} s_0$.
This together with $|\mathcal{S}_0|=s_0$ entails that 
\begin{align*}
|\mathcal{S}_2^c \cap \mathcal{S}_0|
=	|\mathcal{S}_0|-|\mathcal{S}_2 \cap \mathcal{S}_0|
\geq  (1-C_{\ell_1}C_{\lambda}\kappa_n^{-1}) s_0. 
\end{align*}
Observe that $\mathcal{S}_2^c \subset\mbox{supp}(W)$ with asymptotic probability one since by assumption there are no ties in the nonzero components of the Lasso solution in \eqref{eq: hat-b} with asymptotic probability one. Therefore,we have $|\mbox{supp}(W)\cap \mathcal{S}_0|
\geq |\mathcal{S}_2^c \cap \mathcal{S}_0|
\geq  (1-C_{\ell_1}C_{\lambda}\kappa_n^{-1}) s_0$.
This, together with \eqref{eq: bound-S0hat-S0}, yields that
\begin{align}\label{eq: case2-bound}
\frac{|\widehat{\mathcal{S}}_0 \cap \mathcal{S}_0|}{|\mathcal{S}_0|}
\geq \frac{(1-C_{\ell_1}C_{\lambda}\kappa_n^{-1}) s_0-\varphi C_{\ell_1}C_{\lambda}\kappa_n^{-1}s_0}{s_0}
=1-C_{\ell_1}C_{\lambda}(\varphi+1)\kappa_n^{-1}.
\end{align}

Combining \eqref{eq: case1-bound} in Case 1 and \eqref{eq: case2-bound} in Case 2 above, we have shown that with
asymptotic probability one, it holds that
\begin{align*}
|\widehat{\mathcal{S}}_0 \cap \mathcal{S}_0|/|\mathcal{S}_0|
\geq 
1-C_{\ell_1}C_{\lambda}(\varphi+1)\kappa_n^{-1}
\end{align*}
since $\varphi+1=\varphi/(\varphi-1)$ by the definition of $\varphi$. Thus, 
\begin{align*}
\mathrm{power}(\widehat{\mathcal{S}}_0 )=\mathbb{E}\left[|\widehat{\mathcal{S}}_0 \cap \mathcal{S}_0|/|\mathcal{S}_0|\right]
\geq 
1-C_{\ell_1}C_{\lambda}(\varphi+1)\kappa_n^{-1}
\to 1
\end{align*}
as $n\to\infty$, which completes the proof of Theorem~\ref{Theorem2}.

	\section*{Appendix S2: Additional numerical studies}\label{App: S2}

	As suggested by one reviewer, we add one additional simulation study in which the censoring times depend on some of the important covariates and the baseline hazard function $h_0(t)$ is $t$-dependent.

	\indent{\bf Study 3}. In this study, given the covariate vector $\bx_i$, the censoring times $U_i$'s are independently generated from an exponential distribution with mean $c\exp(\bgamma_0^{\top}\bx_i)$, where $\bgamma_0$ is a $p$-dimensional vector and 
	the constant $c$ is chosen to obtain a censoring rate of about $20\%$ and $40\%$, respectively, while the survival time 
	$T_i$'s are generated based on the hazard function
	$h(t|\bx_i) = h_0(t)\exp(\bbeta_0^{\top}\bx_i)$ with $h_0(t)=2t$.   
	We consider three cases for the true coefficient vector $\bbeta_0=(\beta_{01}, \ldots, \beta_{0p})^{\top}$:
	1) Case 1, $\beta_{0j}=5$ for $1\leq j\leq 10$ and $0$; 2) Case 2, $\beta_{0j}=2$ for $1\leq j\leq 10$ and $0$ otherwise; 3) Case 3, $\beta_{0j}=2$ for $1\leq j\leq 20$ and $0$ otherwise.
	We take $\bgamma_0=(\gamma_{01}, \ldots, \gamma_{0p})^{\top}$ with $\gamma_{01}=\gamma_{02}=2$ and $\gamma_{0j}=0$ for $j\geq 3$. Thus, the censoring times only depend on the first two important covariates.  
	We also consider Gaussian distribution and multivariate $t$-distribution for the underlying distribution of the covariate vector $\bx$. For the former, each covariate vector $\bx_i$ is independently generated from a Gaussian distribution $N(\mathbf{0},\bSigma)$ for $i=1, \ldots, n$, where $\bSigma=(\Sigma_{kl})$ with $\Sigma_{kl}=0.5^{|k-l|}$ for $1\leq k, \ell\leq p$. For the latter, each $\bx_i$ is independently generated from the $t$-distribution with mean zero and degrees of freedom $\nu=3$. To be more specific,  
	$\bx_i\thicksim [(\nu-2)/\nu]^{1/2}\u_i/\sqrt{\Gamma_i}$ for $i=1, \ldots, n$,
	where $\u_i\sim N(\textbf{0}, \bSigma)$ with $\bSigma$ given before,
	$\Gamma_i$ is independently sampled from a gamma distribution with the shape and rate parameter both equal to $\nu/2$. We consider different dimensions $(n, p)=(1000, 100)$ and $(n, p)=(400, 1000)$ and set the target FDR level to $q=0.2$.

	\begin{center}
		\begin{table}[h!]%
			\caption{The empirical FDR and power of different methods over 100 replications for Study 3 and target FDR level $q=0.2$.}
			\label{tab-Study3}
			\centering
			\scalebox{0.86}{
			\begin{tabular}{llll cl cl cl}
				\toprule	
				\multirow{2}{*}{\textbf{Distribution}} & \multirow{2}{*}{$(n,p)$} & 
				\multirow{2}{*}{\textbf{CR}} & \multirow{2}{*}{\textbf{Method}} & \multicolumn{2}{c}{\textbf{Case 1}} & \multicolumn{2}{c}{\textbf{Case 2}} & \multicolumn{2}{c}{\textbf{Case 3}}  \\ [-9pt] 
				& & & & \multicolumn{2}{c@{}}{\hrulefill} & \multicolumn{2}{c@{}}{\hrulefill} & \multicolumn{2}{c@{}}{\hrulefill}\\
				&  &  &  & \textbf{FDR} & \textbf{Power}   & \textbf{FDR} & \textbf{Power}  & \textbf{FDR} & \textbf{Power}  \\ 
				\hline
				Gaussian	& $(1000,100)$ & 20$\%$  & CoxLasso     &  0.851  & 1   & 0.805   & 1   & 0.750   & 1 \\
				&              &         & CoxKnockoff  &  0.245  & 1   & 0.222   & 1   & 0.197   & 1 \\
				&              &         & CoxKnockoff+ &  0.183  & 1   & 0.142   & 1   & 0.168   & 1 \\
				&              & 40$\%$  & CoxLasso     &  0.850  & 1   & 0.853   & 1   & 0.762   & 1\\
				&              &         & CoxKnockoff  &  0.224  & 1   & 0.204   & 1   & 0.205   & 1\\
				&              &         & CoxKnockoff+ &  0.159  & 1   & 0.135   & 1   & 0.172   & 1  \\
				\midrule
				Gaussian    & $(400,1000)$ & 20$\%$  & CoxLasso    &  0.959  & 1   & 0.956   & 1   & 0.924   & 1 \\
				&              &         & CoxKnocoff  &  0.235  & 1   & 0.269   & 1   & 0.222   & 1 \\
				&              &         & CoxKnocoff+ &  0.169  & 1   & 0.187   & 1   & 0.180   & 1\\
				&              & 40$\%$  & CoxLasso    &  0.947  & 1   & 0.948   & 1   & 0.898   & 1\\
				&              &         & CoxKnocoff  &  0.222  & 1   & 0.230   & 1   & 0.211   & 1 \\
				&              &         & CoxKnocoff+ &  0.169  & 1   & 0.163   & 1   & 0.180   & 1\\ 			
				\midrule
				$t$         & $(1000,100)$ & 20$\%$  & CoxLasso    &  0.302  & 1   & 0.230 & 0.999 & 0.237 & 1\\
				&              &         & CoxKnocoff  &  0.153  & 1   & 0.114 & 0.999 & 0.120 & 1 \\
				&              &         & CoxKnocoff+ &  0.127  & 1   & 0.102 & 0.999 & 0.110 & 1\\
				&              & 40$\%$  & CoxLasso    &  0.321  & 1   & 0.246 & 0.996 & 0.287 & 0.997 \\
				&              &         & CoxKnocoff  &  0.160  & 1   & 0.124 & 0.996 & 0.147 & 0.996 \\
				&              &         & CoxKnocoff+ &  0.131  & 1   & 0.101 & 0.996 & 0.128 & 0.995\\ 	 
				\midrule
				$t$         & $(400,1000)$ & 20$\%$ & CoxLasso    &  0.721   & 1   & 0.695 & 0.999 & 0.666 & 1 \\
				&              &        & CoxKnocoff  &  0.190   & 1   & 0.185 & 0.999 & 0.154 & 1 \\
				&              &        & CoxKnocoff+ &  0.159   & 1   & 0.153 & 0.999 & 0.134 & 1\\
				&              & 40$\%$ & CoxLasso    &  0.737   & 1   & 0.737 & 1     & 0.743 & 1\\
				&              &        & CoxKnocoff  &  0.170   & 0.999 & 0.171 & 1   & 0.140 & 1\\
				&              &        & CoxKnocoff+ &  0.137   & 0.999 & 0.124 & 1   & 0.124 & 1\\ 			
				\bottomrule
			\end{tabular}}
		\end{table}
	\end{center}

Table~\ref{tab-Study3} reports the empirical FDR and power based on 100 replications for different methods 
under various scenarios in Study 3. As expected, our methods (CoxKnockoff and CoxKnockoff+) are obviously better than CoxLasso in terms of FDR in all settings. To be more specific, CoxKnockoff+ has high power and 
can strictly control the FDR below the target level.  This is consistent with our theory in Section~\ref{sec: theory}. CoxKnockoff also enjoys high power, but has empirical FDR slightly higher than the target level under some scenarios. This is reasonable since the knockoff threshold is designed for controlling the mFDR. However, the CoxLasso method fails to control the FDR although its power is also high.



\begin{thebibliography}{}
	\bibitem[Abramovich et~al., 2006]{abramovich2006adapting}
	Abramovich, F., Benjamini, Y., Donoho, D.~L., \& Johnstone, I.~M. (2006).
	\newblock Adapting to unknown sparsity by controlling the false discovery rate.
	\newblock \textit{The Annals of Statistics}, 34, 584--653.
	
	\bibitem[Antoniadis et~al., 2010]{antoniadis2010dantzig}
	Antoniadis, A., Fryzlewicz, P., \& Letu{\'e}, F. (2010).
	\newblock The {Dantzig} selector in {Cox's} proportional hazards model.
	\newblock \textit{Scandinavian Journal of Statistics}, 37, 531--552.
	
	\bibitem[Barber and Cand{\`e}s, 2015]{barber2015controlling}
	Barber, R.~F. \& Cand{\`e}s, E.~J. (2015).
	\newblock Controlling the false discovery rate via knockoffs.
	\newblock \textit{The Annals of Statistics}, 43, 2055--2085.
	
	\bibitem[Barber and Cand{\`e}s, 2019]{barber2019knockoff}
	Barber, R.~F. \& Cand{\`e}s, E.~J. (2019).
	\newblock A knockoff filter for high-dimensional selective inference.
	\newblock \textit{The Annals of Statistics}, 47, 2504--2537.
	
	\bibitem[Barber et~al., 2020]{barber2020robust}
	Barber, R.~F., Cand{\`e}s, E.~J., \& Samworth, R.~J. (2020).
	\newblock Robust inference with knockoffs.
	\newblock \textit{The Annals of Statistics}, 48, 1409--1431.
	
	\bibitem[Benjamini and Hochberg, 1995]{benjamini1995controlling}
	Benjamini, Y. \& Hochberg, Y. (1995).
	\newblock Controlling the false discovery rate: a practical and powerful
	approach to multiple testing.
	\newblock \textit{Journal of the Royal Statistical Society, Series B},
	57, 289--300.
	
	\bibitem[Benjamini and Yekutieli, 2001]{benjamini2001control}
	Benjamini, Y. \& Yekutieli, D. (2001).
	\newblock The control of the false discovery rate in multiple testing under
	dependency.
	\newblock \textit{The Annals of Statistics}, 29, 1165--1188.
	
	\bibitem[Bradic et~al., 2011]{bradic2011regularization}
	Bradic, J., Fan, J., \& Jiang, J. (2011).
	\newblock Regularization for {Cox’s} proportional hazards model with
	{NP}-dimensionality.
	\newblock \textit{The Annals of Statistics}, 39, 3092--3120.
	
	\bibitem[Breslow, 1974]{breslow1974covariance}
	Breslow, N. (1974).
	\newblock Covariance analysis of censored survival data.
	\newblock \textit{Biometrics}, 30, 89--99.
	
	\bibitem[Cand{\`e}s et~al., 2018]{candes2018panning}
	Cand{\`e}s, E., Fan, Y., Janson, L., \& Lv, J. (2018).
	\newblock Panning for gold: ‘model-x’ knockoffs for high dimensional
	controlled variable selection.
	\newblock \textit{Journal of the Royal Statistical Society, Series B},
	80, 551--577.
	
	\bibitem[Cox, 1972]{cox1972regression}
	Cox, D.~R. (1972).
	\newblock Regression models and life-tables.
	\newblock \textit{Journal of the Royal Statistical Society, Series B},
	34, 187--202.
	
	\bibitem[Dong et~al., 2022]{dong2022reproducible}
	Dong, Y., Li, D., Zheng, Z., \& Zhou, J. (2022).
	\newblock Reproducible feature selection in high-dimensional accelerated
	failure time models.
	\newblock \textit{Statistics $\&$ Probability Letters}, 181, 109275.
	
	\bibitem[Fan et~al., 2012]{fan2012estimating}
	Fan, J., Han, X., \& Gu, W. (2012).
	\newblock Estimating false discovery proportion under arbitrary covariance
	dependence.
	\newblock \textit{Journal of the American Statistical Association},
	107, 1019--1035.
	
	\bibitem[Fan and Li, 2002]{fan2002variable}
	Fan, J. \& Li, R. (2002).
	\newblock Variable selection for {Cox's} proportional hazards model and frailty
	model.
	\newblock \textit{The Annals of Statistics}, 30, 74--99.
	
	\bibitem[Fan et~al., 2020a]{fan2020rank}
	Fan, Y., Demirkaya, E., Li, G., \& Lv, J. (2020a).
	\newblock {RANK}: Large-scale inference with graphical nonlinear knockoffs.
	\newblock \textit{Journal of the American Statistical Association}, 115, 362--379.
	
	\bibitem[Fan et~al., 2020b]{fan2020ipad}
	Fan, Y., Lv, J., Sharifvaghefi, M., \& Uematsu, Y. (2020b).
	\newblock {IPAD}: Stable interpretable forecasting with knockoffs inference.
	\newblock \textit{Journal of the American Statistical Association},
	115, 1822--1834.
	
	\bibitem[Fang et~al., 2017]{fang2017testing}
	Fang, E.~X., Ning, Y., \& Liu, H. (2017).
	\newblock Testing and confidence intervals for high dimensional proportional
	hazards models.
	\newblock \textit{Journal of the Royal Statistical Society, Series B},
	79, 1415--1437.
	
	\bibitem[Huang et~al., 2013]{huang2013oracle}
	Huang, J., Sun, T., Ying, Z., Yu, Y., \& Zhang, C.-H. (2013).
	\newblock Oracle inequalities for the lasso in the {Cox} model.
	\newblock \textit{The Annals of Statistics}, 41, 1142--1165.
	
	\bibitem[Li and Maathuis, 2021]{li2021ggm}
	Li, J. \& Maathuis, M.~H. (2021).
	\newblock {GGM} knockoff filter: False discovery rate control for {Gaussian}
	graphical models.
	\newblock \textit{Journal of the Royal Statistical Society, Series B},
	83, 534--558.
	
	\bibitem[Lin and Lv, 2013]{lin2013high}
	Lin, W. \& Lv, J. (2013).
	\newblock High-dimensional sparse additive hazards regression.
	\newblock \textit{Journal of the American Statistical Association}, 108, 247--264.
	
	
	\bibitem[Liu et~al., 2022]{liu2022model}
	Liu, W., Ke, Y., Liu, J., \& Li, R. (2022).
	\newblock Model-free feature screening and {FDR} control with knockoff features.
	\newblock \textit{Journal of the American Statistical Association}, 117, 428--443.
	
	\bibitem[Liu et~al., 2012]{liu2012variable}
	Liu, X., Peng, Y., Tu, D., \& Liang, H. (2012).
	\newblock Variable selection in semiparametric cure models based on penalized
	likelihood, with application to breast cancer clinical trials.
	\newblock \textit{Statistics in Medicine}, 31, 2882--2891.
	
	\bibitem[Mook et~al., 2009]{mook200970}
	Mook, S., Schmidt, M.~K., Viale, G., Pruneri, G., Eekhout, I., Floore, A., $\ldots$ Van't~Veer, L.~J.
	(2009).
	\newblock The 70-gene prognosis-signature predicts disease outcome in breast
	cancer patients with 1-3 positive lymph nodes in an independent validation
	study.
	\newblock \textit{Breast Cancer Research and Treatment}, 116, 295--302.
	
	\bibitem[Sarkar and Tang, 2022]{sarkar2022adjusting}
	Sarkar, S.~K. \& Tang, C.~Y. (2022).
	\newblock Adjusting the {Benjamini--Hochberg} method for controlling the false
	discovery rate in knockoff-assisted variable selection.
	\newblock \textit{Biometrika}, 109, 1149--1155.
	
	\bibitem[Storey, 2002]{storey2002direct}
	Storey, J.~D. (2002).
	\newblock A direct approach to false discovery rates.
	\newblock \textit{Journal of the Royal Statistical Society, Series B},
	64, 479--498.
	
	\bibitem[Tibshirani, 1997]{tibshirani1997lasso}
	Tibshirani, R. (1997).
	\newblock The lasso method for variable selection in the {Cox} model.
	\newblock \textit{Statistics in Medicine}, 16, 385--395.
	
	\bibitem[van~de Geer and B{\"u}hlmann, 2009]{van2009conditions}
	van~de Geer, S.~A. \& B{\"u}hlmann, P. (2009).
	\newblock On the conditions used to prove oracle results for the lasso.
	\newblock \textit{Electronic Journal of Statistics}, 3, 1360--1392.
	
	\bibitem[Van't~Veer et~al., 2002]{van2002gene}
	Van't~Veer, L.~J., Dai, H., Van De~Vijver, M.~J., He, Y.~D., Hart, A.~A., Mao,
	M., $\ldots$ Friend, S. H. 
	(2002).
	\newblock Gene expression profiling predicts clinical outcome of breast cancer.
	\newblock \textit{Nature}, 415, 530--536.
	
	\bibitem[Weinstein et~al., 2023]{weinstein2020power}
	Weinstein, A., Su, W.~J., Bogdan, M., Barber, R.~F., \& Cand{\`e}s, E.~J.
	(2023).
	\newblock A power analysis for model-x knockoffs with $\ell_p$-regularized
	statistics.
	\newblock \textit{The Annals of Statistics}, to appear.
	
	
	\bibitem[Wu et~al., 2020]{wu2020variable}
	Wu, Q., Zhao, H., Zhu, L., \& Sun, Jianguo (2020).
	\newblock Variable selection for high-dimensional partly linear additive {Cox} model with application to {Alzheimer's} disease.
	\newblock \textit{Statistics in Medicine}, 39, 3120--3134.	
	
	\bibitem[Yu et~al., 2021]{yu2021confidence}
	Yu, Y., Bradic, J., \& Samworth, R.~J. (2021).
	\newblock Confidence intervals for high-dimensional {Cox} models.
	\newblock \textit{Statistica Sinica}, 31, 243--267.
	
	\bibitem[Zhang and Lu, 2007]{zhang2007adaptive}
	Zhang, H.~H. \& Lu, W. (2007).
	\newblock Adaptive lasso for {Cox's} proportional hazards model.
	\newblock \textit{Biometrika}, 94, 691--703.
	
	\bibitem[Zheng et~al., 2018]{zheng2018recovering}
	Zheng, Z., Zhou, J., Guo, X., \& Li, D. (2018).
	\newblock Recovering the graphical structures via knockoffs.
	\newblock \textit{Procedia Computer Science}, 129, 201--207.
	
	
	\bibitem[Zheng et~al., 2022]{zheng2022L0}
	Zheng, Z., Zhang, J., \& Li, Y. (2022).
	\newblock ${L}_0$-Regularized learning for high-dimensional additive hazards regression.
	\newblock \textit{INFORMS Journal on Computing}, 34, 2762--2775.
	
	
	\bibitem[Zhou et~al., 2022]{zhou2022reproducible}
	Zhou, J., Li, Y., Zheng, Z., \& Li, D. (2022).
	\newblock Reproducible learning in large-scale graphical models.
	\newblock \textit{Journal of Multivariate Analysis}, 189, 104934.
	
	\bibitem[Zou, 2008]{zou2008note}
	Zou, H. (2008).
	\newblock A note on path-based variable selection in the penalized proportional
	hazards model.
	\newblock \textit{Biometrika}, 95, 241--247.
\end{thebibliography}

\begin{thebibliography}{}	
	\bibitem[Cand{\`e}s et~al., 2018]{candes2018panning}
	Cand{\`e}s, E., Fan, Y., Janson, L., \& Lv, J. (2018).
	\newblock Panning for gold: ‘model-x’ knockoffs for high dimensional
	controlled variable selection.
	\newblock \textit{Journal of the Royal Statistical Society, Series B},
	80, 551--577.
	
	
	\bibitem[Dong et~al., 2022]{dong2022reproducible}
	Dong, Y., Li, D., Zheng, Z., \& Zhou, J. (2022).
	\newblock Reproducible feature selection in high-dimensional accelerated
	failure time models.
	\newblock \textit{Statistics $\&$ Probability Letters}, 181, 109275.
	
	
	\bibitem[Yu et~al., 2021]{yu2021confidence}
	Yu, Y., Bradic, J., \& Samworth, R.~J. (2021).
	\newblock Confidence intervals for high-dimensional {Cox} models.
	\newblock \textit{Statistica Sinica}, 31, 243--267.	
\end{thebibliography}
\end{document}